\documentclass[12pt]{article}
\usepackage{amsmath}
\usepackage{amstext,amssymb,amsfonts, graphicx,color}

\advance\voffset by -1.5cm
\advance\hoffset by -1.25cm
\def\blue{\color{black}}

\definecolor{darkgreen}{rgb}{0,0.6,0}
\def\thefootnote{\fnsymbol{footnote}}

\def\be{\begin{equation}}
\def\ee{\end{equation}}
\def\ba{\begin{eqnarray}}
\def\ea{\end{eqnarray}}

\newcommand{\N}{{\cal N}}

\newcommand{\CC}{\mathbb{C}}

\newcommand{\nn}{{\nonumber}}

\newcommand{\zb}{{\bar z}}
\newcommand{\xb}{{\bar x}}
\newcommand{\yb}{{\bar y}}

\newcommand{\RR}{{\mathbb R}}

\newcommand{\jhat}{\hat{\jmath}}

\newcommand{\chop}[3]{\mathcal{#1}_{#2}^{(#3)}}
\newcommand{\chops}[3]{\mathcal{#1}_{#2}(#3)}

\newcommand{\kdiff}[4]{\frac{1}{z_{#1#2}}\left(x^2_{#1#2}#3 #4 x_{#1#2}\right)}
\newcommand{\mref}[1]{(\ref{#1})}

\def\<{\langle}
\def\>{\rangle}
\usepackage{srcltx}
\usepackage{color}

\usepackage[english, USenglish]{babel}
\usepackage{epsfig}
\usepackage{slashed}

\begin{document}


\thispagestyle{empty}
\renewcommand{\thefootnote}{\fnsymbol{footnote}}

{\hfill \parbox{2.5cm}{
 DESY 11-111 \\ 
}}

\bigskip\bigskip\bigskip

\begin{center} \noindent \Large \bf
Worldsheet operator product expansions and \\
$p$-point functions in AdS$_3$/CFT$_2$
\end{center}

\bigskip\bigskip\bigskip

\centerline{ \normalsize \bf 
Ingo Kirsch$^{a,}$\footnote[2]{\noindent \tt email: ingo.kirsch@desy.de}
and Tim Wirtz$^{b,}$\footnote[1]{\noindent \tt email: wirtz@tp1.physik.uni-siegen.de}}

\bigskip\bigskip

\centerline{\it ${}^a$ DESY Hamburg, Theory Group,}
\centerline{\it Notkestrasse 85, 22607 Hamburg, Germany}
\vspace{0.3cm}
\centerline{\it ${}^b$ University of Siegen,
Department of Physics,}
\centerline{\it \\ Walter-Flex-Strasse 3,
57068 Siegen, Germany}

\bigskip\bigskip\bigskip

\bigskip\bigskip\bigskip

\renewcommand{\thefootnote}{\arabic{footnote}}

\centerline{\bf Abstract}
\medskip

{We construct the operator product expansions (OPE) of the chiral primary operators
in the worldsheet theory for strings on $AdS_3 \times S^3 \times T^4$. As an interesting
application, we will use the worldsheet OPEs to derive a recursion relation for a
particular class of extremal $p$-point correlators on the sphere. We compare our
result with the corresponding recursion relation previously found
in the symmetric orbifold theory on the boundary of $AdS_3$.}

\newpage
\setcounter{tocdepth}{2}
\tableofcontents

\setcounter{equation}{0}
\section{Introduction}

In the recent years, much progress has been made in matching 
correlation functions in the $AdS_3/CFT_2$ correspondence \cite{Maldacena:1997re}.
In the symmetric product orbifold theory on the boundary of the $AdS_3$ space, two- and
three-point functions of single-cycle twist operators were computed in 
\cite{Jevicki, Lunin1}. In \cite{PRR}, this analysis was extended to
some simple four-point functions, and recursion relations were found for some
extremal $p$-point correlators. In the dual worldsheet theory for string theory
on $AdS_3\times S^3 \times T^4$, the two- and three-point correlators of chiral primary
operators were derived in \cite{Gaberdiel, Pakman1} (see also \cite{Giribet, Cardona2})
and intriguing agreement with the dual boundary correlators was found (Later, 
agreement with supergravity was achieved in \cite{Taylor}, see also \cite{Mihailescu, Arutyunov, Pankiewicz} for earlier work). Recently, in \cite{Cardona}, (the one-particle
contributions of) some extremal four-point correlators  have been computed on the
 worldsheet using a general method
for $SL(2)$ correlation functions developed in \cite{MO}. Again, agreement was found with
the corresponding boundary result of \cite{PRR}. Even though the 
string theory/supergravity and field theory correlators are computed at different points in the moduli space, they must and do agree as predicted by the non-renormalization theorem of
\cite{deBoer}. This theorem states that all three-point functions as well as all
extremal $p$-point functions $(p>3)$ of chiral primary operators are protected along the moduli space \cite{deBoer}.

In this paper we extend the analysis of \cite{Cardona} by deriving a recursion relation
for higher $p$-point correlators in the worldsheet theory. Such $p$-point functions
may be factorized by means of {\em worldsheet} operator product expansions (OPE), which 
should not be mixed up with their dual spacetime OPEs. Their general properties 
were discussed in \cite{Aharony} for string theory on a general $AdS_{d+1}\times W$
background. Here we specialize to $AdS_3\times S^3 \times T^4$ and compute the worldsheet
operator product expansions of chiral primary operators in the associated $H^+_3 \times SU(2)$
Wess-Zumino-Witten (WZW) model. The chiral primaries are
composite operators of the bosonic $H^+_3$ and $SU(2)$ primaries, usually dressed
with some free fermions and ghosts, and their OPEs are obtained by combining the OPEs of the
individual fields. Comparing the thus obtained (unintegrated) worldsheet OPEs with the 
corresponding spacetime OPEs, we find, not surprisingly, a one-to-one realization of the fusion rules of the chiral ring. We will also discuss some structural differences between both kinds of OPEs.

To find the recursion relation, we insert the worldsheet OPEs into a particular class of extremal $p$-point functions 
of chiral primary operators (In an extremal $p$-point function the spacetime scaling of the
$p$-th operator is the sum of the spacetime scalings of the other \mbox{$p-1$} operators). 
After performing the integrals over a (single) worldsheet coordinate and the
$SL(2)$ representation label $h$, in a similar fashion as in \cite{Cardona}, 
the $p$-point function factorizes into the product of a \mbox{$p-1$}-point function
and a three-point function. In this way we find a recursion relation which, up to an overall factor
$F$, is in agreement with the recursion relation of the dual boundary correlator previously found in \cite{PRR}. We will comment on $F$ in the conclusions.

\setcounter{equation}{0}
\section{Worldsheet operator product expansions in $AdS_3$}

In the following we derive the operator product expansions of the chiral primary
operators in the worldsheet theory for string theory on $AdS_3\times S^3 \times T^4$.
In the next section, we will use the resulting OPEs to find a recursion relation
for a particular class of $p$-point functions.

\subsection{Chiral primary operators}\label{sec21}

We begin by summarizing the worldsheet chiral primary operators  
\cite{KLL, Argurio, Pakman1}. Our conventions are as in \cite{Cardona}. 
In particular, it is understood that all operators depend on the complex
worldsheet coordinate $z$, even though we often omit this dependence in
the arguments of the operators.

The worldsheet theory is the product of an $\N = 1$ WZW model on
$H^+_3$, an $\N = 1$ WZW model on $S^3 \simeq SU(2)$ and an $\N = 1$
$U(1)^4$ free superconformal field theory. {\blue We emphasize here that, following
\cite{MO}, we consider an $H^+_3=SL(2,\CC)/SU(2)$ sigma model whose target space is a 
{\em Euclidean} $AdS_3$. Likewise, the dual CFT$_2$ on the boundary is unitary and its time 
variable can be analytically continued to Euclidean time.  In this way
we avoid problems which arise in the definition of operator product expansions
in the (Lorentzian) $SL(2,\RR)$ WZW model \cite{Satoh}--\cite{Fjelstad}.}

The above WZW model has the
affine world-sheet symmetry $\widehat{sl}(2)_k \times
\widehat{su}(2)_{k'} \times u(1)^4$. Criticality of the fermionic string on 
$AdS_3\times S^3$ requires the identification of the levels
$k$ and $k'$ \cite{GKS}, $k=k'$ . The label~$k$ denotes the supersymmetric
level  of the affine Lie algebras and is identified with the 
bosonic levels $k_b$ and $k'_b$ as  $k=k_b-2=k'_b+2$.
 The bosonic currents are $J^a$ for
$SL(2)$ and $K^a$ for $SU(2)$. The free fermions of $SL(2)$ are
denoted by $\psi^a$, those of $SU(2)$ by $\chi^a$ ($a=(+,0,-)$ in
either case). It is convenient to split the bosonic currents 
as 
\begin{align}
J^a = j^a + \jhat^a  \,,\qquad \jhat^a= -\frac{i}{k} 
\varepsilon^a{}_{bc} \psi^a \psi^b \,,
\end{align}
and similarly $K^a$.  Finally the $u(1)^4$ symmetry is described in
terms of free bosons as $i\partial Y^i$, and the corresponding free
fermions are $\lambda_i$ $(i = 1, 2, 3, 4)$.

\medskip
The chiral operators are constructed from the dimension zero operators
\begin{align} \label{Oj}
{\cal O}_j(x,y)= \Phi_h (x) \Phi'_{j}(y)  \qquad{\rm with}\qquad
h=j+1   \,, \quad \textstyle j=0,\frac{1}{2}, ...,\frac{k-2}{2}\,,
\end{align}
where $\Phi_h (x)$ and $\Phi'_{j}(y)$ are the primaries of the bosonic
$H^+_3$ and $SU(2)$ WZW models with dimensions
\begin{align}
\Delta(h) = -\frac{h(h-1)}{k_b-2} \,,\qquad \Delta'(j) = \frac{j(j+1)}{k'_b+2} \,,
\end{align}
respectively.\footnote{As mentioned above, we drop the dependence on the 
worldsheet coordinate $z$. For instance, the holomorphic $H^+_3$ operator is simply denoted by $\Phi_h(x)$ instead of $\Phi_h(x,z)$.} The labels $x$ and $y$ correspond to
the $SL(2)$ and $SU(2)$ representation labels $m$ and $m'$, respectively.  Our
conventions for these models can be found in appendix~A of 
\cite{Cardona}.
Since $h=j+1$, the operators ${\cal O}_j(x,y)$ have vanishing
conformal dimensions, $\Delta(h)+\Delta'(j)=0$.

\subsubsection*{Neveu-Schwarz sector}

In the {\em Neveu-Schwarz} sector there are two families of chiral primaries. In the
$-1$ picture they are\footnote{In \cite{KLL}, these operators are denoted 
by ${\cal W}^-_j$ and ${\cal X}^+_j$, respectively.}
\begin{align}
 {\cal O}^{(0)}_j(x,y) &= e^{-\phi} \psi(x) {\cal O}_j(x,y) \,,  
 \label{pminusone}\\
 {\cal O}^{(2)}_j(x,y) &= e^{-\phi} \chi(y) {\cal O}_j(x,y) \,, 
\end{align}
where the fields $\psi(x)$ and $\chi(y)$ are given by
\begin{align}\label{psichi}
\psi(x)&=-\psi^+ + 2x \psi^3 - x^2 \psi^-\,, \nn\\
\chi(y)&=-\chi^+ + 2y \chi^3 + y^2 \chi^-\,. 
\end{align}
The bosonized superghost field $e^{-\phi}$ ensures that the operators
have ghost number $-1$.

Sometimes we will also need the corresponding ghost number $0$
operators, which are obtained from (\ref{pminusone}) by acting with
the picture changing operator $\Gamma_{+1}$. These operators will be
needed to get the correct ghost number in the correlators.
The ghost number $0$ operators are \cite{Pakman1,
  Gaberdiel}
\begin{align}
 \tilde {\cal O}^{(0)}_j(x,y) &= \left((1 - h) \jhat(x) + j(x) + 
 \textstyle\frac{2}{k} \psi(x) \chi_a P^a_y \right) {\cal O}_j(x, y)\,,\label{pzero}\\
 \tilde {\cal O}^{(2)}_j(x,y) &= \left( h \hat k(y) + k(y) + \textstyle\frac{2}{k}
 \chi(y) \psi_A D^A_x\right) {\cal O}_j(x, y) \,,
\end{align}
where the operators $D^A_x$ and $P^a_y$ are
\begin{align}
D_x^- = \partial_x \,, \quad D_x^3 = x \partial_x + h \,,
\quad D_x^+ = x^2 \partial_x + 2h x \,,\nn\\
P_y^- = - \partial_y \,, \quad P_y^3 = y \partial_y - j \,,
\quad P_y^+ = y^2 \partial_y - 2j y \,.
\end{align}
Here we used again the compact notation 
\begin{align}
\jhat(x) &= -\hat \jmath^+ + 2x \hat \jmath^3 - x^2\hat \jmath^- \,,\nn\\
\hat k(y) &= -\hat k^+ + 2y \hat k^3 + y^2\hat k^- \,, \quad
etc.
\end{align}

\subsubsection*{Ramond sector}

In the {\em Ramond sector} there are also two families of chiral primaries,
${\cal O}^{(a)}_j(x,y)$ with $a=\pm 1$. For their construction we need
the spin operators
\begin{align}
S_{[\varepsilon_1, \varepsilon_2, \varepsilon_3]}
= e^{\frac{i}{2}(\varepsilon_1 \hat H_1 +\varepsilon_2 \hat H_2 
  +\varepsilon_3 \hat H_3)} \,,
\end{align} 
where $\varepsilon_I=\pm 1$ and $\hat H_i$ ($i=1,2,3$) are bosonized fermions
related to $\psi^a$ and $\chi^a$ ($a=\pm,0$), as in \cite{Pakman1} (Similarly, $\hat H_{4,5}$
are related to the fermions on the $T^4$, $\lambda^i$ ($i=1,2,3,4$) \cite{Pakman1}). Then, in the $-1/2$ and $-3/2$ picture the chiral primaries are given 
by\footnote{${\cal O}^{(+1)}_j$ and ${\cal O}^{(-1)}_j$ contain $s^{1}_-$
and $s^{2}_-$, respectively.
In \cite{KLL}, these operators are denoted by ${\cal Y}^\pm_j$.}
\begin{align}
{\cal O}^{(a)}_j(x,y) &= e^{-\frac{\phi}{2}} s^{1,2}_-(x,y) {\cal O}_j(x,y) 
\qquad (a=\pm 1) \,,
\end{align}
and
\begin{align}
\tilde {\cal O}^{(a)}_j(x,y) &= -\sqrt{k} (2h-1)^{-1} 
e^{- \frac{3\phi}{2}} s^{1,2}_+(x,y) {\cal O}_j(x,y) \,,
\end{align}
respectively, where
\begin{align}
s^1_\pm (x, y) &= S_\pm(x,y) e^{+ \frac{i}{2}(\hat H_4 - \hat H_5)} \,,\qquad  
s^2_\pm (x, y) = S_\pm(x,y) e^{- \frac{i}{2}(\hat H_4 - \hat H_5)}
\end{align}
and 
\begin{align}
S_\pm(x,y)= \mp xy i S_{[--\pm]} \mp x S_{[-+\mp]}
+ y i S_{[+-\mp]} + S_{[++\pm]} \,.
\end{align}

\subsubsection*{Full chiral primary operators}

The full chiral primary operators are given by the product of a
holomorphic with an anti-holomorphic operator,
\begin{align}\label{fullchiral}
{\cal O}^{(A,\bar A)}_{j}(x,\bar{x}, y,\bar{y})
\equiv {\cal O}^{(A)}_{j}(x, y) {\bar {\cal O}}^{(\bar A)}_{j}(\bar x, \bar 
y) \,,
\end{align}
where $A = 0, a, 2$ and $\bar A = \bar 0, \bar a, \bar 2$. When integrated
over the worldsheet, these operators are dual to the chiral primary operators 
$O^{(A, \bar A)}_{n}$ ($n$-cycle twist operators with $n=2j+1$) in the symmetric
orbifold theory on the boundary of $AdS_3$, defined {\em e.g.}\ in \cite{Jevicki, Pakman1, PRR}.

\subsection{Worldsheet operator product expansions}\label{secAK}

The general structure of worldsheet operator product expansions for
strings on \mbox{$AdS_{d+1} \times W$} was studied in \cite{Aharony}. The
vertex operators of this theory ${\mathbb O}_{h,j}$ are usually 
labeled by the spacetime scaling dimension $h$ associated with the spacetime conformal 
group $SO(d + 1, 1)$ and a collective label $j$ denoting some internal quantum numbers. 
Let us restrict to $d=2$. 
As exemplified in section~\ref{sec21}, for the special case of
$AdS_3 \times S^3 \times T^4$, the vertex
operators are products of the primaries $\Phi_h(z, x)$ and $\Phi'_j(z, y)$ 
of the bosonic $H^+_3$ and $SU(2)$ WZW models, dressed by a polynomial in
the bosonic and fermionic worldsheet fields and their 
derivatives~\cite{KLL, Argurio, GKS}.  
These operators depend on both the worldsheet coordinate
$z$ as well as the $SL(2)$ and $SU(2)$ representation labels $x$ and $y$. As argued
in \cite{deBoer:1998pp}, the label $x$ can be identified with the coordinate on the boundary. Moreover, the  $SL(2)$ current algebra on the string worldsheet induces a 
Virasoro algebra in spacetime conformal field theory.
In addition to the usual worldsheet conformal weight $\Delta=\Delta(h,j)$
the vertex operators therefore also have a spacetime scaling dimension related to 
$h$.\footnote{The exact spacetime scaling
depends on the actual form of the operator, {\em e.g.}\ $h[{\cal O}^{(0)}_j]=
h[{\cal O}_j]+h[\psi]=h-1$ for the operator ${\cal O}^{(0)}_j$ defined in 
(\ref{pminusone}).} Physical vertex operators have worldsheet dimension $\Delta(h,j)=1$.

The Hilbert space of the worldsheet theory contains only the normalizable
vertex operators with $h=\frac{1}{2}+is$ ($s\in \RR$). For such operators,
the most general form of an $AdS_3$ worldsheet OPE is, in the limit $z \rightarrow 0$, 
\cite{Aharony}:
\begin{align}\label{OPEAharony}
{\mathbb O}_{1}(0) {\mathbb O}_{2}(x,\bar{x},z,\bar{z})
&= \sum_j \int_{{\cal C}} dh \int d^2x'\, \frac{|z|^{2 (\Delta(h,j)-\Delta(1)-\Delta(2))}}{|x|^{\alpha}|x'|^{\beta}
|x'-x|^{\gamma}} {\cal F}(j_i,j, h_i,h) {\mathbb O}_{h,j}(x',\bar x',0,0)\nonumber\\
 &~~~+ \textmd{descendants} \,,
\end{align}
where ${\cal F}$ is related to the 2-point and 3-point functions on the worldsheet.
The parameters $\alpha$, $\beta$ and $\gamma$ are functions of the spacetime conformal weights
of the operators \mbox{${\mathbb O}_{i}\equiv 
{\mathbb O}_{h_i,j_i}$ $(i=1,2)$} and ${\mathbb O}_{h,j}$, respectively. $\Delta(1)$, $\Delta(2)$
and $\Delta(h,j)$ denote the corresponding
worldsheet conformal weights. 
The dependence on $z$ and $x$ is completely determined by conformal invariance.
The OPE contains an integral over the contour $h=\frac{1}{2}+is$, which is denoted by ${\cal C}$.
In the following, we ignore contributions coming from the worldsheet descendants. 

The above OPE is not directly applicable to 
worldsheet operators which are dual to spacetime operators. Such operators
are non-normalizable and therefore not part of the Hilbert space. Instead they 
have spacetime scalings related to $h$ located on the real axis of the 
complex $h$-plane. The OPE of such non-normalizable operators is obtained by careful
analytic continuation in $h$. As shown in \cite{Aharony}, this amounts to
the inclusion of additional discrete contributions from the poles of ${\cal F}$. Otherwise, the form of (\ref{OPEAharony}) is preserved.

\subsection{Worldsheet operator product expansions of chiral primary operators}

We now compute the OPE (\ref{OPEAharony}) for the case that the worldsheet operators
are chiral primary. We begin by constructing the OPE of the dimension-zero 
operators
\begin{align}
\chops{O}{j}{x,\xb,y,\yb}= \Phi_{h}(x,\xb)\Phi'_{j}(y,\yb)  \qquad (h=j+1)\,,
\end{align}
which form an essential part of the chiral primaries, as discussed after
(\ref{Oj}). The OPE is obtained from the OPEs of the $H^+_3$ and $SU(2)$ 
fields $\Phi_{h}(x,\xb)$ and $\Phi'_{j}(y,\yb)$.

The OPE of two $H^+_3$ primaries was found in \cite{Teschner1999}.
As shown in Appendix~\ref{appB}, it can be written as
\begin{align}
&\Phi_{h_2}(x_2,\xb_2) \Phi_{h_1}(x_1,\xb_1) 
= \int_{{\cal C^+}} dh \frac{C(h_1,h_2,h)|z_{12}|^{-2\Delta_{12}}|x_{12}|^{-2h_{12}}}{B(h)}\Phi_{h}(x_{1},
\xb_1) \label{sl2ope}
\,,
\end{align}
with $h_{12}=h_1+h_2-h$ and $\Delta_{12}=\Delta_1+\Delta_2-\Delta$ $({\cal C}^+=1/2+i\RR^+)$. $C(h_1,h_2,h_3)$ and $B(h)$ are the $SL(2)$ structure
constants and the scaling of the $SL(2)$ two-point function, respectively. 
Similarly, the OPE of two $SU(2)$ primaries is given by~\cite{Zamolodchikov, Dotsenko}
\begin{align}
& \Phi_{j_2}'(y_2,\yb_2) \Phi_{j_1}'(y_1,\yb_1) 
=\sum_{j} C'(j_1,j_2,j)|z_{12}|^{-2\Delta'_{12}}|y_{12}|^{2j_{12}}\Phi_{j}'(y_1,\yb_1) \,,
\end{align}
with $j_{12}=j_1+j_2-j$ and $\Delta_{12}=\Delta'_1+\Delta'_2-\Delta'$.
In both OPEs we ignored the contribution from current algebra descendants.
Combining both OPEs yields the operator product expansion
\begin{align}
&\chops{O}{j_2}{x_2,\xb_2,y_2,\yb_2} \chops{O}{j_1}{x_1,\xb_1,y_1,\yb_1}\nn\\
&~~~= \sum_j \int_{\cal C^+} dh  \frac{C'C|z_{12}|^{-2(\Delta_{12} + \Delta'_{12})}|y_{12}|^{2j_{12}}}{B(h)|x_{12}|^{2h_{12}}}\Phi_{h}(x_{1},\xb_1,z_1,\zb_1)\Phi_{j}'(y_1,\yb_1,z_1,\zb_1)\nn\\
&~~~=\sum_j \int_{\cal C^+} dh  \frac{C'C|z_{12}|^{2(\Delta(h) + \Delta'(j))}|y_{12}|^{2j_{12}}}{B(h)|x_{12}|^{2h_{12}}}\chops{O}{j,h}
{x_1,\xb_1,y_1,\yb_1} \label{equ:zerochiralprimary} \,.
\end{align}
In the last line we defined the more general operators ${\cal O}_{j,h} \equiv \Phi_h\Phi'_j$, 
for which the labels $h$ and $j$ are not related in any way. Recall that the 
resulting operator ${\cal O}_{j,h}$ need not be physical.

\medskip

Let us now construct the OPE of the operator $\mathcal{O}_{j}^{(0,0)}$ in the
$-1$~picture and $\tilde{\mathcal{O}}_{j}^{(0,0)}$ in the $0$~picture, which 
are defined by (\ref{pminusone}) and (\ref{pzero}), respectively.
We start from the expression
\begin{align}
&\tilde{\mathcal{O}}_{j_2}^{(0,0)}(x_2,\bar x_2, y_2,\bar y_2)
\mathcal{O}_{j_1}^{(0,0)}(x_1,\bar x_1,y_1,\bar y_1) \\ 
&~~~= \big((1-h_2)\hat{\jmath}(x_2) + j(x_2) + \textstyle\frac{2}{k}\psi(x_2)\chi_a P^a_{y_2}\big) e^{-\phi}\psi(x_1) \nn\\
&~~~\times\big((1-h_2)\bar{\hat{\jmath}}(\bar x_2) + \bar{j}(\bar{x}_2) + \textstyle\frac{2}{k}\bar \psi(\bar x_2)\bar \chi_a 
P^a_{\bar y_2}\big) e^{-\bar \phi}
\bar\psi(\bar x_1) \chops{O}{j_2}{x_2,\xb_2,y_2,\yb_2} \chops{O}{j_1}{x_1,\xb_1,y_1,\yb_1}\,.\nn
\end{align}
Using the OPEs \mref{equ:singularjphi}-\mref{equ:singularpsipsi} in appendix~\ref{appC} and
the identity
\begin{align}\label{chipi}
2 \chi_a P^a_y=\chi(y)\partial_y-j\partial_y\chi(y)\, ,
\end{align}
this can also be written as
\begin{align}\label{224}
&\tilde{\mathcal{O}}_{j_2}^{(0,0)}(x_2,\bar x_2, y_2,\bar y_2)
\mathcal{O}_{j_1}^{(0,0)}(x_1,\bar x_1,y_1,\bar y_1) \nn\\ 
&~~~=\left\vert(1-h_2)\left(\mathcal{D}_{21}^{(-1)}\psi(x_1)\right) + \psi(x_1)\mathcal{D}_{21}^{(h_1)}+ \frac{x_{21}^2}{z_{21}}\left(\chi(y_{2})\partial_{y_{2}} -j_{2}\partial_{y_{2}}\chi(y_{2})\right)\right\vert^2\nn\\ &~~~~~~\times e^{-\phi-\bar\phi }
\chops{O}{j_2}{x_2,\xb_2,y_2,\yb_2} \chops{O}{j_1}{x_1,\xb_1,y_1,\yb_1} \nn\\
&~~~= \left\vert\frac{x_{21}}{z_{21}}\psi(x_1)\left( (1-h_2) 2 
+ x_{21} \partial_{x_1}  - 2 h_1 \right) + \frac{x_{21}^2}{z_{21}} \chi(y_{2})
\partial_{y_2}+... \right\vert^2 e^{-\phi-\bar \phi }\mathcal{O}_{j_2}\mathcal{O}_{j_1} \,.
\end{align} 
The ellipses denote further terms involving derivatives of the type
$\partial \psi$ and $\partial \chi$. In this analysis we neglect descendants
and therefore ignore such terms. In the following we will also need to Taylor expand $\chi(y_{2})
= \chi(y_{1})+y_{12} \partial \chi(y_{1})+...$ and again drop derivatives of~$\chi$.   $|...|^2$~indicates that there is the same factor
in anti-holomorphic variables. 

Substituting (\ref{equ:zerochiralprimary}) into (\ref{224}),
we evaluate the derivatives on ${\cal O}_j$ such that 
$x_{21}\partial_{x_1}\rightarrow h_{12}$ and $y_{21} \partial_{y_2} \rightarrow j_{12}$
under the integral. We obtain
\begin{align}
&\tilde{\mathcal{O}}_{j_2}^{(0,0)}(x_2,\bar x_2, y_2,\bar y_2)
\mathcal{O}_{j_1}^{(0,0)}(x_1,\bar x_1,y_1,\bar y_1) \nn\\ 
&~~~= \sum_j \int_{\cal C^+} dh 
\frac{C'C|z_{12}|^{2(\Delta(h) + \Delta'(j)-1)}|y_{12}|^{2j_{12}}}{B(h)|x_{12}|^{2(h_{12}-1)}}
 \left( (h_1+h_2+h-2)^2
{\cal{O}}_{j,h}^{(0,0)}
(x_1,\xb_1,y_1,\yb_1) \right. \nn\\&\qquad\qquad ~~~~~~\left. +\, {
(j_{12})^2 \frac{ |x_{21}|^{2}}{|y_{21}|^{2}}{\cal{O}}_{j,h}^{(2,2)}
(x_1,\xb_1,y_1,\yb_1)} + ... \right) \label{fullOPE}\,,
\end{align}
where we ignored possible terms involving 'mixed' operators
of the type ${\cal{O}}_{j,h}^{(0,2)}$ and ${\cal{O}}_{j,h}^{(2,0)}$.

\medskip
Before we continue, let us recall how the fusion rules in the conformal field theory on
the boundary can be reproduced from the worldsheet description \cite{Pakman1}. 
The operators in the OPE must obey $U(1)$ charge conservation
(as measured by the $SU(2)$ generator $K_0^3$, see \cite{Pakman1}). Chiral (anti-chiral) operators in the boundary
CFT are mapped to highest (lowest) weight states of $SU(2)$ in the worldsheet theory, {\em i.e.}\ 
$M = J$ ($M = -J$). $U(1)$ charge conservation in the fusion of two worldsheet operators,
symbolically
\begin{align}
{\cal O}^{(*)}_{j_1} \times {\cal O}^{(*)}_{j_2} = [ {\cal O}^{(*)}_{j_3} ] \,,
\end{align}
therefore requires \cite{Pakman1}
\begin{align}
J=J_1+J_2\,, \label{J}
\end{align}
where $J_i=j_i+ a_i$ and $a_i=0,1/2,1$ for the holomorphic
operators ${\cal O}^{(0)},
{\cal O}^{(a)},{\cal O}^{(2)}$, respectively. The fusion of two $SU(2)$ primary states
requires $j_3 \leq j_1+j_2$ and therefore (\ref{J}) implies
\begin{align} \label{a}
a_3 \geq a_1+a_2 \,.
\end{align} Clearly, the fusion rules must also obey the spin-statistics relations ${\rm NS} \times {\rm NS} \rightarrow {\rm NS}$, 
${\rm NS} \times {\rm R} \rightarrow {\rm R}$, ${\rm R} \times {\rm NS} \rightarrow {\rm R}$, and ${\rm R} \times {\rm R} \rightarrow {\rm NS}$,
where ${\rm NS}$ and ${\rm R}$ refer to the operators in the Neveu-Schwarz sector 
(${\cal O}^{(0)},{\cal O}^{(2)}$) and Ramond sector (${\cal O}^{(a)}$), respectively.  
This allows for the following fusion rules in the holomorphic sector:
\begin{align}
(0) \times (0) &= (0)+(2) \,,\nn\\
(0) \times (2) &= (2) \,,\nn\\
(0) \times (a) &= (a) \,,\nn\\
(a) \times (a) &= (2) \,. \label{fusionhol}
\end{align}
Similar fusion rules hold in the anti-holomorphic sector. The four cases (\ref{fusionhol}) can be freely combined between holomorphic and anti-holomorphic operators. Note however that in the fusion $(0,0) \times (0,0) \rightarrow (0,0)
+ (2,2)$ the resulting operator must be the same in the holomorphic and anti-holomorphic sector, {\em i.e.}\ the combinations $(0,2)$ and $(2,0)$ do not appear \cite{Pakman1}. In (\ref{fullOPE}) the fusion rules therefore only allow for terms involving the operators
\begin{align}
{\cal{O}}_{j,h}^{(0,0)}: \qquad &j=j_1+j_2 \equiv \tilde j\,, \nonumber\\
{\cal{O}}_{j,h}^{(2,2)}: \qquad &j=j_1+j_2-1 \equiv \tilde j-1 \,, \label{jvals}
\end{align}
where the $j$-values have been determined using (\ref{J}).
Terms proportional to ${\cal{O}}_{j,h}^{(0,2)}$ and ${\cal{O}}_{j,h}^{(2,0)}$
are forbidden by the worldsheet fusion rules.

\medskip

In order to compare the worldsheet OPE with the corresponding boundary OPE, 
we need to rescale the operators as in \cite{Cardona} such that their 
(integrated) two-point functions scale as unity. For instance,
the operators ${\cal O}^{(0,\bar 0)}_{j}(x,\bar{x})$ will be rescaled 
as
\begin{align}
{\mathbb O}^{(0, 0)}_{j}(x,\bar{x}) &=
 {\frac{\sqrt{2\pi^2}}{\sqrt{k\,B(h)(2h-1)}}}
g_s\,{\cal O}^{(0,\bar 0)}_{j}(x,\bar{x}) \, .
\end{align}
Then, as shown in detail in appendix \ref{secrescaling}, the OPE of the rescaled
operators ${{\mathbb O}}_{j}^{(0,\bar{0}) }$ following from (\ref{fullOPE}) is
\begin{align}
&\tilde{{\mathbb O}}_{j_2}^{(0,\bar{0}) }(x_2,\bar{x}_2;y_2,\bar{y}_2){\mathbb O}_{j_1}^{(0,\bar{0}) }(x_1,\bar{x}_1;y_1,\bar{y}_1)\label{equ:exactope} \\
&~~~= \int_{\cal C^+} dh~\frac{2h-1}{2\pi^2k} \frac{|z_{21}|^{2(\Delta(h)-1)}}
{|x_{21}|^{2(h_{21}-1)}} \left( 
|z_{21}|^{2\Delta'(\tilde j)}\,{\mathbb G}^{(000)}_3(j_1,j_2,\tilde j,h)
{\mathbb O}_{\tilde j,h}^{(0,\bar{0}) }(x_1,\bar{x}_1;y_1,\bar{y}_1) \right. \nn\\
&~~~~~~~~~~~~~~~\qquad\quad\left. 
+ \,  {|x_{21}|^{2}} |z_{21}|^{2\Delta'(\tilde j-1)}\,{\mathbb G}^{(002)}_3(j_1,j_2,\tilde j-1,h) 
{\mathbb O}_{\tilde j-1,h}^{(2,{2}) }(x_1,\bar{x}_1;y_1,\bar{y}_1)\right) \,, \nonumber
\end{align}
where in the last line we defined the coefficients 
\begin{align}\label{G3}
{\mathbb G}^{(000)}_3(j_1,j_2, j_3,h_3)
&\equiv P(j_1,j_2,j_3,h_3) \,
 \frac{g_s}{k} \frac{(h_1 +h_2+h_3-2)^2}{\prod_i (2h_i-1)^\frac{1}{2}}\,,\nn\\
{\mathbb G}^{(002)}_3(j_1,j_2, j_3,h_3)
&\equiv P(j_1,j_2,j_3,h_3) \,
\frac{g_s}{k} \frac{(j_{1}+j_2-j_3)^2}{\prod_i (2h_i-1)^\frac{1}{2}} \,,
\end{align}
and 
\begin{align}
P(j_1,j_2,j_3,h_3) \equiv \frac{{C}{C}'~2\pi}{\sqrt{B(h_1)B(h_2)B(h_3)~c_{\nu}}} 
\qquad (c_{\nu} = 1/(2\pi^4k^3))\,.
\end{align}
The factor $P(j_1,j_2,j_3,h_3)$ reflects the fact that $h_3$ is not
related to $j_3$ in the third operator. This factor would be just one, $P(j_1,j_2,j_3,h_3)=1$, 
if $h_3$ were
related to $j_3$ by $h_3=j_3+1$.\footnote{This can be seen by using the identity
(4.29) in \cite{Cardona}. This identity has first been found in
\cite{Gaberdiel,Pakman1}.} In that case, and if $j_3$ is related to $j_{1}+j_2$ as
in (\ref{jvals}),  the coefficients reduce to the 
extremal three-point correlators
\begin{align}
{\mathbb G}^{(000)}_3(j_1,j_2,j_3,h_3) \vert_{h_3= j_3+1} = 
\left<{\mathbb O}^{(0, 0)}_{j_1}(\infty){\mathbb O}^{(0,
      0)}_{j_2}(1)\tilde{{\mathbb O}}^{(0, 0)}_{j_3}(0) \right> \,,\nn\\
{\mathbb G}^{(002)}_3(j_1,j_2, j_3,h_3)\vert_{h_3= j_3+1} = 
 \left<{\mathbb O}^{(0, 0)}_{j_1}(\infty) \tilde{{\mathbb O}}^{(0,
    0)}_{j_2}(1){\mathbb O}^{(2, 2)}_{j_3}(0) \right>      \,, \label{ex}
\end{align}
found in \cite{Pakman1, Gaberdiel}. Note, for instance, that 
the $U(1)$ charge conservation $j_3=j_1+j_2$
is equivalent to $h_3=h_1+h_2-1$, if $h_3=j_3+1$. However, we stress that we do {\em not}
assume any relation between $h$ and $\tilde j$ at this stage, {\em i.e.}\ 
the operators on the right-hand-side of (\ref{equ:exactope}) need not be physical.

\medskip
The other OPEs allowed by the fusion rules are computed in a similar way. We find 
\begin{align}
&\tilde{{\mathbb O}}_{j_2}^{(0,\bar{0}) }(x_2,\bar{x}_2;y_2,\bar{y}_2){\mathbb O}_{j_1}^{(2,\bar{2}) }(x_1,\bar{x}_1;y_1,\bar{y}_1)\nn\\
&~~~= \int_{\cal C^+} dh~\frac{2h-1}{2\pi^2k} \frac{|z_{21}|^{2(\Delta(h)+\Delta'(\tilde j)-1)}}
{|x_{21}|^{2(h_{21}-1)}}  \,{\mathbb G}^{(022)}_3(j_1,j_2,\tilde j,h)
{\mathbb O}_{\tilde j,h}^{(2,\bar{2}) }(x_1,\bar{x}_1;y_1,\bar{y}_1)  \,, \label{OPE2}\\
&\tilde{{\mathbb O}}_{j_2}^{(0,\bar{0}) }(x_2,\bar{x}_2;y_2,\bar{y}_2){\mathbb O}_{j_1}^{(a,\bar{a}) }(x_1,\bar{x}_1;y_1,\bar{y}_1)\nn\\
&~~~= \int_{\cal C^+} dh~\frac{2h-1}{2\pi^2k} \frac{|z_{21}|^{2(\Delta(h)+\Delta'(\tilde j)-1)}}
{|x_{21}|^{2(h_{21}-1)}}  \,{\mathbb G}^{(0aa)}_3(j_1,j_2,\tilde j,h)
{\mathbb O}_{\tilde j,h}^{(a,\bar{a}) }(x_1,\bar{x}_1;y_1,\bar{y}_1)  \,, \label{OPE3}\\
&{{\mathbb O}}_{j_2}^{(a,\bar{a}) }(x_2,\bar{x}_2;y_2,\bar{y}_2){\mathbb O}_{j_1}^{(b,\bar{b}) }(x_1,\bar{x}_1;y_1,\bar{y}_1)\nn\\
&~~~= \int_{\cal C^+} dh~\frac{2h-1}{2\pi^2k} \frac{|z_{21}|^{2(\Delta(h)+\Delta'(\tilde j)-1)}}
{|x_{21}|^{2(h_{21}-1)}}  \,{\mathbb G}^{(ab2)}_3(j_1,j_2,\tilde j,h)
{\mathbb O}_{\tilde j,h}^{(2,\bar{2}) }(x_1,\bar{x}_1;y_1,\bar{y}_1)  \,,\label{OPE4}
\end{align} 
with
\begin{align}
{\mathbb G}^{(022)}_3(j_1,j_2, j_3,h_3)
&\equiv P(j_1,j_2,j_3,h_3) 
 \frac{g_s}{k} \frac{(-h_1 +h_2+h_3)^2}{\prod_i (2h_i-1)^\frac{1}{2}}\,, \label{G022}\\
 {\mathbb G}^{(0aa)}_3(j_1,j_2, j_3,h_3)
&\equiv P(j_1,j_2,j_3,h_3) 
 \frac{g_s}{k} 
 { \frac{(h_1+h_2+h_3-2)^2}{(2h_3-1)^2}} \frac{(2h_1-1)^\frac{1}{2}(2h_3-1)^\frac{1}{2} }{(2h_2-1)^\frac{1}{2}}
 \,, \label{G0aa}\\ 
 {\mathbb G}^{(ab2)}_3(j_1,j_2, j_3,h_3)
&\equiv P(j_1,j_2,j_3,h_3)  \frac{g_s}{k} \frac{(2h_1-1)^\frac{1}{2}(2h_2-1)^\frac{1}{2} }{(2h_3-1)^\frac{1}{2}} \delta^{ab}
 \,.\label{Gab2}
\end{align}
The correlators (\ref{G022})--(\ref{Gab2})  
reduce again to the extremal three-point functions computed in \cite{Pakman1},
if $h_3=j_3+1$. Note that the total ghost number is preserved in the OPEs.

\subsection{Discussion and comparison with boundary operator product expansions}\label{secdisc}

Some comments on the worldsheet operator product expansions (\ref{equ:exactope})
are in order. Similar statements will hold for the OPEs (\ref{OPE2})--(\ref{OPE4}).

First, let us first compare (\ref{equ:exactope}) with the general form (\ref{OPEAharony}).
Defining $\Delta(h,j)\equiv\Delta(h)+\Delta'( j)+1$, which is the worldsheet conformal 
dimension of  ${\mathbb O}_{j,h}^{(0, 0)}$ (and ${\mathbb O}_{j,h}^{(2, 2)}$), we find that at small $z$ and small $x$ (\ref{equ:exactope}) agrees with the general form (\ref{OPEAharony}),
since the chiral primaries have conformal dimension $\Delta(1)=\Delta(2)=1$ and $|z_{21}|^{2(\Delta(h,j)-\Delta(1)-\Delta(2))}=|z_{21}|^{2(\Delta(h)+\Delta'(j)-1)}$.
Recall also that ${\mathbb O}_{j,h}^{(0, 0)}$ and ${\mathbb O}_{j,h}^{(2, 2)}$ scale
differently in $x$, $h^{(0)}=h-1$ and $h^{(2)}=h$ \cite{Cardona}. The total $x$-dependence
should be $|x_{21}|^{2(h^{(A)}-h_1^{(0)}-h_2^{(0)})}$ with $A=0,2$ in the first and second term of  (\ref{equ:exactope}), respectively. Therefore there is an additional factor $|x_{21}|^2$ 
in the second term of (\ref{equ:exactope}).  Consequently, we find that the OPE has
the correct scaling in both $x$ and $z$. (In (\ref{equ:exactope}) we have already used
$U(1)$ charge conservation such that there is no sum over $j$ anymore).
 
Second, another peculiar feature of (\ref{equ:exactope}) is the appearance of the factor
\begin{align} 
\frac{2h-1}{2\pi^2k} \,.
\end{align}
As we will see later, when we use the OPE inside a general correlator, this factor will
cancel against the residue of the $h$-integral, which is proportional to the
inverse of the derivative of the $SL(2)$ conformal weight,
$(\partial_h \Delta)^{-1} \propto k/(2h-1)$.

Third, it is also interesting to compare the worldsheet OPE (\ref{equ:exactope}) with the corresponding 
spacetime OPE of $n$-cycle twist operators of the type $O^{(0,0)}_{n}$ which are dual to 
the worldsheet operators ${\mathbb O}_{j}^{(0, 0)}$. This OPE is given by \cite{PRR}\footnote{See \cite{PRR} for a precise definition of the operators $O^{(0,0)}_{n}$ and the corresponding OPE.} 
\begin{align} \label{stope}
O^{(0,0)}_{n_2} O^{(0,0)}_{n_1} = C_3 O^{(0,0)}_{\tilde n} 
+ C'_3 O^{(2,2)}_{\tilde n-2} + ... \,,
\end{align}
with $\tilde n=n_1+n_2-1$ and structure constants $C_3$ and $C_3'$. The 
ellipses indicate terms coming from multi-cycle operators.
Given that the cycle lengths $n_i$ are related to $j_i$ by $n_i=2j_i+1$ 
(and $\tilde n=2\tilde j+1$), we observe a structural resemblance between the worldsheet and the 
spacetime OPE, cf.\ (\ref{equ:exactope}) with (\ref{stope}). In particular, both 
OPEs satisfy the fusion relation $(0,0) \times (0,0) \rightarrow (0,0) +
(2,2)$ of the chiral-chiral ring. More general, we find that the worldsheet OPEs
(\ref{equ:exactope}), (\ref{OPE2})-(\ref{OPE4}) mimic the fusion rules of
the $(c, c)$ ring in the spacetime conformal field theory, 
\begin{align}
(0, 0) \times (0, 0) &= (0, 0) + (2, 2)\,,\nn\\
(0, 0) \times (2, 2) &= (2, 2) \,,\nn\\
(0, 0) \times (a, a) &= (a, a) \,,\nn\\
(a, a) \times (a, a) &= (2, 2) \,. \label{fusion}
\end{align}
In fact, upon integration over the worldsheet coordinates, the worldsheet OPE 
(\ref{equ:exactope}) becomes identical to the spacetime OPE (\ref{stope}) 
(multi-cycle contributions ignored).\footnote{This can be seen by setting $z_1=0$ and
$z=z_2$ and performing the integral over $z$ and $h$ as described in  
section~3 below. After the integration over $h$, ${\mathbb G}^{(000)}_3$ and ${\mathbb G}^{(002)}_3$ 
have reduced to the extremal correlators (\ref{ex}) which are identical to the coefficients
$C_3$ and $C_3'$ appearing in (\ref{stope}) \cite{Pakman1, Gaberdiel}.}  

Fourth, one might worry that (\ref{equ:exactope})
still depends on the spacetime coordinates $x$, while the spacetime 
OPE (\ref{stope}) has no singularities.
We will see however in the next section that, when the OPE is employed inside an extremal $p$-point 
correlator of chiral primary operators, the $x$-dependence will drop out 
(Basically the integration over $h$ will yield a relation between $h$
and $\tilde j$ which eliminates the $x$-dependence
in both terms in  (\ref{equ:exactope}).).

\setcounter{equation}{0}
\section{Recursion relation for worldsheet $p$-point functions}

In this section we derive a recursion relation for a particular extremal 
worldsheet $p$-point function and compare it with the corresponding relation for
the dual boundary correlator previously computed in~\cite{PRR}. 

A simple worldsheet $p$-point function on the sphere is given by the product of $p$ 
(rescaled) operators 
${\mathbb O}_{j}\equiv {\mathbb O}^{(0, 0)}_{j}$,
\begin{align} \label{defGp}
  {\mathbb G}_p &\equiv{\mathbb G}_p^{j_1,...,j_p}
= g_s^{-2}  \left\<{\tilde{\mathbb O}}_{j_p}(\infty)
  {\mathbb O}_{j_{p-1}}(1) \left( \prod_{i=2}^{p-2} \int d^2z_i \,
   \tilde{{\mathbb O}}_{j_i}(x_i,\bar x_i;z_i,\bar z_i) \right)
   {\mathbb O}_{j_1}(0)\right\> , 
\end{align}
with the extremality condition
\begin{align}\label{extr}
j_p=\sum_{i=1}^{p-1} j_i \,.
\end{align}
Modular invariance has been used to fix three of the $p$ worldsheet points as
$z_{1,p-1,p}=0,1,\infty$. Similarly, 
the continuous $SL(2)$ representation labels are chosen as $x_{1,p-1,p}=0,1,\infty$.
The $x$ labels will later be identified with the complex
coordinates in the spacetime conformal field theory \cite{deBoer:1998pp}.   
The correlator ${\mathbb G}_p$ involves $p-2$ ghost number zero and $2$ ghost number
$-1$ operators, $\tilde{{\mathbb O}}^{(0, 0)}_{j}$ and ${\mathbb O}^{(0,
  0)}_{j}$, respectively.  Recall that the total ghost number of a
correlator on a genus-$g$ surface must be $-\chi=-(2-2g)$, which is
$-2$ on the sphere.

\medskip
We now show that the $p$-point functions ${\mathbb G}_p$ satisfy the
recursion relation
\begin{align}\label{OPEG}
{\mathbb G}_p &\simeq  \left\< {\mathbb O}_{\tilde j}^{(0,0)}(\infty) \tilde{\mathbb O}^{(0,0)}_{j_{2}}(1) 
{\mathbb O}^{(0,0)}_{j_{1}}(0) \right\> {\mathbb G}_{p-1}  
\end{align} 
with $\tilde j=j_1+j_2$. The symbol $\simeq$ indicates
that (\ref{OPEG}) is true up to a factor $F$ which currently cannot be reproduced
on the worldsheet. This factor is coming from two-particle contributions in the
intermediate channel, which are nonlocal on the worldsheet. The factor $F$
has however been determined in the dual symmetric orbifold theory. The recursion
relation for the dual boundary correlators $C_p$ is given by 
\begin{align} \label{bdyrecursion}
C_p  = \frac{n_p}{\tilde n} \left\< {O}^{(0,0)\dagger}_{\tilde n}(\infty) {O}^{(0,0)}_{n_{2}}(1) 
{O}^{(0,0)}_{n_{1}}(0) \right\> C_{p-1} 
\end{align}
with $\tilde n=n_1+n_2-1$ \cite{PRR}. The non-renormalization theorem of \cite{deBoer} predicts
the equivalence of both recursion relations such that $F$ can be identified as 
 $F=\frac{n_p}{\tilde n}=\frac{2j_p+1}{2\tilde j+1}$.

\medskip
Proof of (\ref{OPEG}): Substituting the worldsheet OPE (\ref{equ:exactope}) 
into ${\mathbb G}_p$, we obtain\footnote{Within a $p$-point function the
integration over the half-axis ${\cal C}^+= 1/2 + i\RR^+$ can be extended to an integration over
the full axis ${\cal C} = 1/2 + i\RR$ \cite{Teschner1999}.}
\begin{align}
  {\mathbb G}_p &= g_s^{-2}
   \int d^2z_2   \,\int_{\cal C} dh  \left\<{\tilde{\mathbb O}}_{j_p}(\infty)
  {\mathbb O}_{j_{p-1}}(1) \left( \prod_{i=3}^{p-2} \int d^2z_i \,
   \tilde{{\mathbb O}}_{j_i}(x_i,\bar x_i;z_i,\bar z_i) \right)
   {\mathbb O}_{\tilde j,h}(0) \right\>  \nn \\
&~~~\times \frac{2h-1}{2\pi^2k} \frac{|z_2|^{2(\Delta(h)+\Delta'(\tilde  j)-1)}}{|x_2|^{2(h_{21}-1)}} \,{\mathbb G}^{(000)}_3(j_1,j_2,\tilde j,h) + ... \nn\\
& =   \int d^2z \, \int_{\cal C} dh
\frac{2h-1}{2\pi^2k} \frac{|z|^{2(\Delta(h)+\Delta'(\tilde j)-1)}}{|x|^{2(h_{21}-1)}} 
 {\mathbb G}_{p-1} \,{\mathbb G}^{(000)}_3(j_1,j_2,\tilde j,h)+...
 \,,\label{integrand}
\end{align}
where we set $z=z_2$ ($x=x_2$) and introduced the short hand notation ${\mathbb G}_{p-1}$ for
${\mathbb G}_{p-1}^{j,j_3,...,j_p}$. The ellipses indicate that there is in principle a second
contribution from the operator ${\mathbb O}^{(2,2)}_{\tilde j-1, h}$ in the OPE (\ref{equ:exactope}). This contribution is zero, as will be shown below.

The integrals over $z$ and $h$ can be done as in the case of four-point functions 
\cite{Cardona, MO}. As in \cite{Cardona}, we need to do the $z$-integral before
the $h$-integral. In that case we have to be careful about the occurrence of 
divergencies and regularize the $z$-integral by introducing a cutoff $\varepsilon$ \cite{MO}.
Later, after the integrations, we will eventually take the limit $\varepsilon\rightarrow 0$.
In general it is not known how to compute the $z$-integral over the whole 
range of $z$, but it can be computed in the limit of small
$|z|<\varepsilon$. In this region, the $z$-integral can be performed by elementary methods, 
\begin{align} \label{I1}
 \int_{|z|<\varepsilon} d^2z\,|z|^{2(\lambda(h)-1)} 
=&\frac{\pi}{\lambda(h)} \varepsilon^{2\lambda(h)}
\end{align} 
with $\lambda(h)=\Delta(h)+\Delta'(\tilde j)$. As discussed in \cite{Aharony, MO}, the
integral only over $|z|<\varepsilon$ captures the single-cycle (or, in higher dimensions, 
single-trace) terms in the spacetime OPE. By performing the
integral only over $|z|<\varepsilon$, we omit nonlocal contributions from
the large $z$ region, which are expected to give the double-cycle terms in the
spacetime OPE \cite{MO, Aharony}. This limitation prevents us from deriving
the overall factor $F$, which is known to arise from double-cycle operators 
in the spacetime OPE \cite{PRR}.

{\blue
We now turn to the integration over $h$. In general, after the $z$ integration, there are 
additional discrete contributions coming from poles in the integrand of (\ref{integrand})
\cite{MO, Aharony}.
Such contributions arise when the poles cross the integration contour during 
\begin{itemize}
\item[i)] the analytic continuation in $j_1$ and $j_2$ (or $h_{1,2}=j_{1,2}+1$), and 
\item[ii)] the shift of the contour from $h = 1/2+is$ to $h = h_0+is$ ($s \in \RR$), 
where $h_0$ is defined by $\lambda(h_0)=0$.\footnote{It is convenient to shift the contour in this
way since, as we will see, most of the pole contributions vanish during the shift.}
\end{itemize}
There are altogether four types of poles \cite{MO, Aharony}:
\begin{align}
\textmd{type I:}\qquad &\lambda=0 \,, \nn\\
\textmd{type II:}\qquad &h = h_1 + h_2 + n \,,\nn\\
\textmd{type III:}\qquad & h = k-h_1-h_2+n \,,\nn\\
\textmd{type IV:}\qquad & h = |h_1 - h_2| - n \,, 
\qquad n \in \{0,1,2,...\}\,.\nn
\end{align} 
The poles of type II-IV are poles in the structure constants
$C(h,h_1,h_2)$. As discussed extensively in \cite{Aharony}, none of
these poles contributes to the integral, at least if the preceding
$z$ integration is restricted to the regime $|z|<\varepsilon$. Even though naively one might
interpret the contributions from the poles of type II as
``double-cycle'' operators in the spacetime CFT, such contributions
go to zero in the $\varepsilon \rightarrow 0$ limit \cite{Aharony} (This is
in agreement with the general expectation \cite{Aharony} that contributions from double-cycle
operators arise non-locally, {\em i.e.} at large $z$ and not in the $|z|<\varepsilon$ region). 
Type~III poles do not appear if one assumes $h_1+h_2<\frac{k+1}{2}$ \cite{MO}. Other than
the poles of type II, the type IV poles may contribute both during the analytic
continuation and the additional shift in the contour. It was found
in~\cite{Aharony} that the contribution coming from crossing the 
contour during the analytic continuation is exactly the opposite of
that during the subsequent shift of the contour. In effect, the
poles of type IV do not modify the final result.

We are left with poles of type I, $\lambda(h)=0$, corresponding
to $h=h_0\equiv\tilde j+1$.} The residue of this pole is 
\begin{align}
{\rm Res}(f;h_0)=\frac{\pi \varepsilon^{2\lambda(h_0)}}{\lambda'(h_0)}
\, \frac{2h_0-1}{2\pi^2 k} {\mathbb G}_{p-1} \,{\mathbb G}^{(000)}_3 \,,
\end{align}
where $f$ is the integrand of (\ref{integrand}) and $'\equiv\partial_h$.
Remarkably, the first and second factor on the right-hand side cancel
each other (up to $2\pi$), since $\lambda'(h_0)=\partial_h \Delta(h_0)$. Moreover, the
$x$-dependence drops out since $h_{21}-1=h_2+h_1-h_0-1=j_2+j_1-\tilde j=0$.
Applying the residue theorem, we thus obtain
\begin{align}\label{35}
 {\mathbb G}_{p} &=  
 {\mathbb G}_{p-1} \,{\mathbb G}^{(000)}_3(j_1,j_2,\tilde j,h=\tilde j+1) \nn\\
 &= \left\< {\mathbb O}_{\tilde j}^{(0,0)}(\infty) \tilde{\mathbb O}^{(0,0)}_{j_{2}}(1) 
{\mathbb O}^{(0,0)}_{j_{1}}(0) \right\> {\mathbb G}_{p-1}  \,,
\end{align}
which is nothing but (\ref{OPEG}).

\medskip
We still have to show that in (\ref{integrand}) there are no contributions from the 
operator~${\mathbb O}^{(2,2)}_{\tilde j-1, h}$. The additional 
term in the integrand of (\ref{integrand}) is proportional to 
\begin{align}
\frac{|z|^{2(\Delta(h)+\Delta'(\tilde  j-1)-1)}}{|x|^{2(h_{21}-2)}} 
\,{\mathbb G}^{(002)}_3(j_1,j_2,\tilde j-1,h)
\end{align}
and has a pole at $h=\tilde j$. After applying the residue theorem, the $x$-dependence drops out,
since $|x|^{2(h_2+h_1-h-2)}=|x|^{2((j_2+1)+(j_1+1)-\tilde j-2)}=1$
and we get the additional contribution
\begin{align}\label{2ndcontribution}
\left\< {\mathbb O}_{\tilde j-1}^{(2,2)}(\infty) \tilde{\mathbb O}^{(0,0)}_{j_{2}}(1) 
{\mathbb O}^{(0,0)}_{j_{1}}(0) \right\> {\mathbb G}'_{p-1} \,,
\end{align}
where ${\mathbb G}'_{p-1}$ is defined by
\begin{align}
{\mathbb G}_{p-1}^{\prime\,\tilde j-1, j_3, ..., j_p} = g_s^{-2} 
   \left\< {\mathbb O}_{j_4}^{(0,0)}(\infty) \tilde{\mathbb O}^{(0,0)}_{j_3}(1) 
X {\mathbb O}_{\tilde j-1}^{(2,2)}(0) \right\> 
\end{align}
and $X$ denotes the product of $p-4$ $\tilde {\mathbb O}_j^{(0,0)}$ operators.

Clearly, for $p=4$, the three-point correlator 
${\mathbb G}_{3}^{\prime\,\tilde j-1, j_3, j_4}$ is zero, as can be seen as follows. 
The extremality condition (\ref{extr}) for $G_4^{j_1,j_2,j_3,j_4}$ can be written as  
\begin{align}
j_4=j_1+j_2+j_3= \tilde j + j_3 \,,
\end{align}
which is formally the $U(1)$ charge conservation for the
fusion of ${\mathbb O}_{\tilde j-1}^{(2,2)}$ and $\tilde{\mathbb O}^{(0,0)}_{j_3}$.
However, the fusion rules require $a_4 \geq \tilde{a} + a_3$ (cf.~with (\ref{a})),
which is violated since $a_4=0$ and $\tilde a+a_3=1+0=1$, implying ${\mathbb G}'_{3}{}^{\tilde j-1,j_3,j_4}=0$.
A similar argument holds for $p>4$.  Thus, the term (\ref{2ndcontribution}) vanishes identically.

\setcounter{equation}{0}
\section{Conclusions}

In this paper we studied the worldsheet realization of  
the chiral ring structure of the $N=(4,4)$ 
symmetric orbifold theory on the boundary of  $AdS_3 \times S^3 \times T^4$. 
Our main results are the (unintegrated)
{\em worldsheet} operator product expansions (\ref{equ:exactope}) and (\ref{OPE2})--(\ref{OPE4}),
which nicely reflect the fusion rules of the chiral ring. Despite the similarity 
to the dual spacetime OPEs, there are also some structural differences which 
we discussed at length in section~\ref{secdisc}.
In particular, 
the worldsheet OPEs are not simply given by the (extremal) worldsheet three-point functions of chiral primary operators \cite{Gaberdiel, Pakman1}, as
one might naively expect. In fact, the operators ${\mathbb O}_{h,j}$ appearing on the right hand side of the worldsheet OPEs need not even be physical, {\em i.e.}\ there is a priori no relation between the $SL(2)$ and $SU(2)$ labels $h$ and $j$, whereas $h=j+1$ for chiral primaries. In this respect, the OPEs are  more general than the three-point functions in \cite{Gaberdiel, Pakman1}. 
However, when the worldsheet OPEs are integrated over the worldsheet coordinates,
the $h$ integral turns out to have a pole at $h=j+1$, and the worldsheet OPEs 
become identical to those of the spacetime CFT.

As an interesting application, 
we used the worldsheet OPEs to derive a recursion relation for a particular class of
extremal $p$-point correlators on the worldsheet. Our result (\ref{OPEG}) 
for the correlator (\ref{defGp}) agrees with the recursion relation 
for the dual boundary $p$-point function \cite{PRR}, up to a simple overall factor $F=n_p/\tilde n$. 
In the spacetime OPE
the factor $F$ comes from two-cycle operators, whose contributions are not
suppressed in extremal correlators at large~$N$. Unfortunately, these
contributions arise nonlocally on the worldsheet and are presently
not very well understood \cite{Aharony}. It would be highly desirable to
understand in more detail how multi-cycle (or, in general, multi-trace) operators 
are treated in worldsheet OPEs.

In this paper (and its precursors \cite{Gaberdiel}--\cite{Cardona2},\cite{Cardona})
worldsheet $p$-point functions on $AdS_3 \times S^3$ (with NSNS fluxes) 
are computed on the {\em full quantum} level. This may be compared
to the {\em semi-classical} treatment of worldsheet $p$-point functions 
for string theory on $AdS_5 \times S^5$ (with RR fluxes), see
e.g.~\cite{Zarembo}--\cite{Escobedo}.  To gain more insight into the latter
approach, it would be interesting to repeat such semi-classical computations on $AdS_3\times S^3$ 
and compare the results with the already known quantum correlators. It may also be of interest to attempt a full quantum computation on $AdS_3$ backgrounds with Ramond-Ramond fluxes, perhaps
using techniques suggested in \cite{Troost}.

\section*{Acknowledgment}

We would like to thank Carlos Cardona, Matthias Gaberdiel, Volker Schomerus 
and J\"org Teschner for useful discussions and comments on the paper.
Part of this work was done during DESY's summer student programme 2010. 
T.~W.\ thanks DESY for its hospitality during the summer school.

\section*{Appendix}
\appendix

\setcounter{equation}{0}
\section{OPE of $H^+_3$ primaries}\label{appB}

In the following we derive the worldsheet operator product expansion of 
chiral primary operators in the $H^+_3$ model. -- Important note: Other than in the 
rest of the paper, we use the conventions of Teschner \cite{Teschner1999} 
in this appendix, {\em i.e.}\ we use $j$ to label the $H^+_3$ states.

The worldsheet OPE of two $H^+_3$ primaries is \cite{Teschner1999}\footnote{We
interchange the labels $1\leftrightarrow 2$. In the following 
we ignore the contribution from descendants.} 
\begin{align}
&\Phi_{j_1}(x_1,\xb_1,z_1, \zb_1)\Phi_{j_2}(x_2,\xb_2,z_2,\zb_2)\nn\\
&~~~=\int_{{\cal C}^+} dj_3\, C(j_1,j_2,j_3) |z_{12}|^{-2\Delta_{12}} 
({\cal J}_{12}(j_3) \Phi_{-j_3-1}) (z_2,\zb_2)\,, \label{SLOPE1}
\end{align}
where
\begin{align}
({\cal J}_{12}(j_3) \Phi_{-j_3-1}) (z_2,\zb_2) \equiv \int_{\CC} d^2x_3\,
|x_{12}|^{2j_{12}} |x_{23}|^{2j_{23}} |x_{31}|^{2j_{31}} \Phi_{-j_3-1}(x_3,\xb_3,z_2,\zb_2)\,.
\end{align}
Here $\Delta_{12}=\Delta_1+\Delta_2-\Delta_3$, $j_{12}=j_1+j_2-j_3$, etc. We prefer
to express the OPE in terms of $\Phi_{j_3}$ rather than $\Phi_{-j_3-1}$. We therefore
substitute the expression
\begin{align}
({\cal J}_{12}(j_3) \Phi_{-j_3-1}) (z_2,\zb_2) = \frac{\gamma(-2j_3)}{(-\pi) \gamma(-j_{23})\gamma(-j_{31})}
\frac{1}{B(j_3)} ({\cal J}_{12}(-j_3-1) \Phi_{j_3}) (z_2,\zb_2)
\end{align}
into (\ref{SLOPE1}) and obtain 
\begin{align}
&\Phi_{j_1}(x_1,\xb_1,z_1, \zb_1)\Phi_{j_2}(x_2,\xb_2,z_2,\zb_2) \\
&~~~=\int_{{\cal C}^+} dj_3\, C(j_1,j_2,j_3) |z_{12}|^{-2\Delta_{12}} 
\frac{\gamma(-2j_3)}{(-\pi) \gamma(-j_{23})\gamma(-j_{31})}
\frac{1}{B(j_3)}\nn\\
&~~~~~~\times 
\int_{\CC} d^2x_3\,
|x_{12}|^{-2(-j_1-j_2-j_3-1)} |x_{23}|^{-2(1+j_{31})} |x_{31}|^{-2(1+j_{23})} \Phi_{j_3}(x_3,\xb_3,z_2,\zb_2) \nn\,.
\end{align}
We now simplify the expression by computing the $x_3$-integral
\begin{align}
I&=\int_{\CC} d^2 t'\,
|t|^{-2(-j_1-j_2-j_3-1)} |t'|^{-2(1+j_{31})} |t-t'|^{-2(1+j_{23})} \Phi_{j_3}({ x_2-}t',
{\bar x_2-}\bar t',z_2,\zb_2) \,,
\end{align}
where we have defined $t=x_{12}$ and $t'=x_{23}$. Denoting $t=|t|\hat t$ and defining
$y=t'/|t|$, we get
\begin{align}
I&=|t|^{-2(-j_1-j_2+j_3+1)} \int_{\CC} d^2 t'\,
 (|t'|/|t|)^{-2(1+j_{31})} |(t'/|t|-\hat t)|^{-2(1+j_{23})} \Phi_{j_3}({ x_2-}t',{ \bar x_2-}\bar t',z_2,\zb_2) \nn\\
 &=|t|^{2j_{12}} \int_{\CC} d^2 y \,
 |y|^{-2(1+j_{31})} |y-\hat t|^{-2(1+j_{23})} \Phi_{j_3}({ x_2-}y|t|,{ \bar x_2-}\bar y|t|,z_2,\zb_2) 
  \,.
\end{align}
In the OPE, $x_1$ and $x_2$ are assumed to be close to each other such that
$|t|$ is small. {\blue We also ignore the subleading contributions
from space-time descendants.} We may then Taylor expanded the operator 
$\Phi_{j_3}(x_2-y|t|,\bar x_2-\bar y|t|,z_2,\zb_2)$ around $x_2$  and 
obtain\footnote{An almost identical expansion was done in Eq.~(2.10)
in \cite{Aharony}.}
\begin{align}
I 
 &\approx |t|^{2j_{12}} \Phi_{j_3}(x_2,\xb_2,z_2,\zb_2) \int_{\CC} d^2 y \,
 |y|^{-2(1+j_{31})} |y-\hat t|^{-2(1+j_{23})} \,.
\end{align}
Using the identity
\begin{align}
\int_{\mathbb{C}} d^2y \,|y|^{2a}|1-y|^{2b} = -\pi\frac{\gamma(-1-a-b)}{\gamma(-a)\gamma(-b)}\,,
\end{align}
the integral $I$ becomes 
\begin{align}
I&= (-\pi) |x_{12}|^{2j_{12}} 
\frac{\gamma(1+2j_3)}{\gamma(1+j_{31})\gamma(1+j_{23})}\Phi_{j_3}(x_2,\xb_2,z_2,\zb_2)\,.
\end{align}
Thus,
\begin{align}
&\Phi_{j_1}(x_1,\xb_1,z_1, \zb_1)\Phi_{j_2}(x_2,\xb_2,z_2,\zb_2) \nn \\
&~~~=\int_{{\cal C}^+} dj_3\, C(j_1,j_2,j_3) |z_{12}|^{-2\Delta_{12}} 
\frac{1}{B(j_3)} |x_{12}|^{2j_{12}} \Phi_{j_3}(x_2,\xb_2,z_2,\zb_2) \,.
\end{align}
Replacing $j \rightarrow -h$ $(\Phi_j \rightarrow \Phi_h)$, we get
(\ref{sl2ope}).

\setcounter{equation}{0}
\section{Some correlators and operator product expansions}\label{appC}

In this appendix we list some
worldsheet operator product expansions used in section~2.
It is convenient to express these OPEs in terms of the operator
\begin{align}
\mathcal{D}^{(h_i)}_{ki}&=\kdiff{k}{i}{\partial_{x_i}}{-2 h_i}\,,
\end{align}
where $h_i$ denotes the spacetime scaling of the operator it
acts on. Some important worldsheet operator product expansions 
are \cite{Pakman1, Cardona}:
\begin{align}
j(x_k) \Phi_{h_i}(x_i)
    &\sim {\cal D}_{ki}^{(h_i)} \Phi_{h_i}(x_i) \,, \label{equ:singularjphi}\\
j(x_1) j(x_2) &\sim (k+2) \frac{x_{12}^2}{z_{12}^2} +{\cal D}_{12}^{(-1)}
j(x_2) \,, \label{equ:singularjj}\\
\jhat(x_1) \jhat(x_2) &\sim -2 \frac{x_{12}^2}{z_{12}^2} +{\cal D}_{12}^{(-1)}
\jhat(x_2) \,,\label{equ:singularhjhj} \\
\hat \jmath(x_1) \psi(x_2) &\sim  {\cal D}_{12}^{(-1)}
\psi(x_2) \,,\label{equ:singularjpsi}\\
\psi(x_1)\psi(x_2) &\sim k\frac{x_{12}^2}{z_{12}} \,. \label{equ:singularpsipsi}
\end{align}

\setcounter{equation}{0}
\section{Rescaling the operators in the OPE} \label{secrescaling}

In this appendix we compute the rescaled OPE (\ref{equ:exactope}). 
For comparison with the boundary theory, it is useful to rescale the 
operators such that, when integrated over $z$, their two-point functions 
are just one (integration over $z_{1,2}$). The
rescaled operators are \cite{Cardona} 
\begin{align}
{\mathbb O}^{(0, 0)}_{j}(x,\bar{x}) &=
 {\frac{\sqrt{2\pi^2}}{\sqrt{k\,B(h)(2h-1)}}}
g_s\,{\cal O}^{(0,\bar 0)}_{j}(x,\bar{x}) \, ,\nonumber\\
{\mathbb O}^{(a,\bar a)}_{j}(x,\bar x)&=\sqrt{\frac{2\pi^2(2h-1)}{B(h)}}g_s
{\cal O}^{(a,\bar a)}_{j}(x,\bar x) \label{renorma}\, .  
\end{align}
The operator ${\cal O}^{(2, 2)}_{j}(x,\bar{x})$ is rescaled as
${\cal O}^{(0, 0)}_{j}(x,\bar{x})$ (Tilded operators are rescaled as their 
untilded partners). 
Then, substituting the OPE (\ref{fullOPE}) into 
\begin{align}
&\tilde{{\mathbb O}}_{j_2}^{(0,\bar{0}) }(x_2,\bar{x}_2;y_2,\bar{y}_2){\mathbb O}_{j_1}^{(0,\bar{0}) }(x_1,\bar{x}_1;y_1,\bar{y}_1)\nn\\
&~~~=\frac{2\pi^2 g_s^2}{k\sqrt{B(h_1)(2h_1-1)B(h_2)(2h_2-1)}} \chop{\tilde O}{j_2}{0,\bar{0}}(x_2,\bar{x}_2;y_2,\bar{y}_2)\chop{O}{j_1}{0,\bar{0}}(x_1,\bar{x}_1;y_1,\bar{y}_1) \,,
\end{align}
we get
\begin{align}
&\tilde{{\mathbb O}}_{j_2}^{(0,\bar{0}) }(x_2,\bar{x}_2;y_2,\bar{y}_2){\mathbb O}_{j_1}^{(0,\bar{0}) }(x_1,\bar{x}_1;y_1,\bar{y}_1) \\
&~~~=\sum_j \int_{\cal C} dh \frac{|z_{12}|^{2(\Delta(h)+\Delta'(j)-1)}|y_{12}|^{2j_{12}}}{|x_{12}|^{2(h_{12}-1)}}\frac{(2h-1)}{\sqrt{(2h-1)(2h_2-1)(2h_1-1)}} \frac{g_s\sqrt{2\pi^2}~{C}'{C}}{\sqrt{k~B(h_1)B(h_2)B(h)}} 
  \nn\\&~~~~~~\times
\left( (h_1+h_2+h-2)^2{\mathbb O}_{j,h}^{(0,\bar{0}) }(x_1,\bar{x}_1;y_1,\bar{y}_1)+
  (j_{12})^2 \frac{|x_{21}|^{2}}{|y_{21}|^{2}}
{\mathbb O}_{j,h}^{(2,\bar{2}) }(x_1,\bar{x}_1;y_1,\bar{y}_1)\right)  \nn
 \,, 
\end{align}
which can be written as (\ref{equ:exactope}).



\begin{thebibliography}{99}

\bibitem{Maldacena:1997re}
  J.~M.~Maldacena,
  {\it The large N limit of superconformal field theories and supergravity},
  Adv.\ Theor.\ Math.\ Phys.\  {\bf 2}, 231 (1998)
  [Int.\ J.\ Theor.\ Phys.\  {\bf 38}, 1113 (1999)]
  {\tt [arXiv:hep-th/9711200]}.

\bibitem{Jevicki}
  A.~Jevicki, M.~Mihailescu and S.~Ramgoolam,
  {\it Gravity from CFT on $S^N(X)$: Symmetries and interactions},
  Nucl.\ Phys.\  B {\bf 577}, 47 (2000)
  {\tt [arXiv:hep-th/9907144]}.


\bibitem{Lunin1}
  O.~Lunin and S.~D.~Mathur,
  {\it Correlation functions for M(N)/S(N) orbifolds},
  Commun.\ Math.\ Phys.\  {\bf 219}, 399 (2001)
  {\tt [arXiv:hep-th/0006196]};
  O.~Lunin and S.~D.~Mathur,
  {\it Three-point functions for M(N)/S(N) orbifolds with {$N = 4$} supersymmetry},
  Commun.\ Math.\ Phys.\  {\bf 227}, 385 (2002)
  {\tt [arXiv:hep-th/0103169]}.


\bibitem{PRR}
  A.~Pakman, L.~Rastelli and S.~S.~Razamat,
  {\it Extremal Correlators and Hurwitz Numbers in Symmetric Product Orbifolds},
  Phys.\ Rev.\  D {\bf 80}, 086009 (2009)
  {\tt [arXiv:0905.3451 [hep-th]]}.


\bibitem{Gaberdiel}
  M.~R.~Gaberdiel and I.~Kirsch,
  {\it Worldsheet correlators in AdS(3)/CFT(2)},
  JHEP {\bf 0704}, 050 (2007)
  {\tt [arXiv:hep-th/0703001]}.

\bibitem{Pakman1}
  A.~Dabholkar and A.~Pakman,
  {\it Exact chiral ring of AdS(3)/CFT(2)},
  Adv.\ Theor.\ Math.\ Phys.\  {\bf 13}, 409 (2009)
  {\tt [arXiv:hep-th/0703022]};
  A.~Pakman and A.~Sever,
  {\it Exact N=4 correlators of AdS(3)/CFT(2)},
  Phys.\ Lett.\  B {\bf 652}, 60 (2007)
  {\tt [arXiv:0704.3040 [hep-th]]}.

\bibitem{Giribet}
  G.~Giribet, A.~Pakman and L.~Rastelli,
  {\it Spectral Flow in AdS(3)/CFT(2)},
  JHEP {\bf 0806}, 013 (2008)
  {\tt [arXiv:0712.3046 [hep-th]]}.

\bibitem{Cardona2}
  C.~A.~Cardona and C.~A.~Nunez,
  {\it Three-point functions in superstring theory on $AdS_3\times S^3 \times T^  4$}, JHEP {\bf 0906}, 009 (2009)
  {\tt [arXiv:0903.2001 [hep-th]]}.

\bibitem{Taylor}
  M.~Taylor,
  {\it Matching of correlators in $AdS_3/CFT_2$},
  JHEP {\bf 0806}, 010 (2008)
  {\tt [arXiv:0709.1838 [hep-th]]}.
  
\bibitem{Mihailescu}
  M.~Mihailescu,
  {\it Correlation functions for chiral primaries in D = 6 supergravity on  
  AdS(3) x S(3)},
  JHEP {\bf 0002}, 007 (2000)
  {\tt [arXiv:hep-th/9910111]}.

\bibitem{Arutyunov}
  G.~Arutyunov, A.~Pankiewicz and S.~Theisen,
  {\it Cubic couplings in D = 6 N = 4b supergravity on AdS(3) x S(3)},
  Phys.\ Rev.\  D {\bf 63}, 044024 (2001)
  {\tt [arXiv:hep-th/0007061]}.

\bibitem{Pankiewicz}
A.~Pankiewicz, {\it
Six-dimensional supergravities and the AdS/CFT correspondence},
Diploma Thesis, University of Munich, October 2000.

\bibitem{Cardona}
  C.~A.~Cardona and I.~Kirsch,
  {\it Worldsheet four-point functions in AdS(3)/CFT(2)},
  JHEP {\bf 1101}, 015 (2011)
  {\tt [arXiv:1007.2720 [hep-th]]}.


\bibitem{MO}
  J.~M.~Maldacena and H.~Ooguri,
  {\it Strings in AdS(3) and the SL(2,R) WZW model. III: Correlation  functions},
  Phys.\ Rev.\  D {\bf 65}, 106006 (2002)
  {\tt [arXiv:hep-th/0111180]}.


\bibitem{deBoer}
  J.~de Boer, J.~Manschot, K.~Papadodimas and E.~Verlinde,
  {\it The chiral ring of AdS3/CFT2 and the attractor mechanism},
  JHEP {\bf 0903}, 030 (2009)
  {\tt [arXiv:} {\tt 0809.0507 [hep-th]]}.
  
\bibitem{Aharony}
  O.~Aharony and Z.~Komargodski,
  {\it The space-time operator product expansion in string theory duals of 
  field theories},
  JHEP {\bf 0801}, 064 (2008)
  {\tt [arXiv:0711.1174 [hep-th]]}.


\bibitem{KLL}
  D.~Kutasov, F.~Larsen and R.~G.~Leigh,
  {\it String theory in magnetic monopole backgrounds},
  Nucl.\ Phys.\  B {\bf 550}, 183 (1999)
  {\tt [arXiv:hep-th/9812027]}.

\bibitem{Argurio}
  R.~Argurio, A.~Giveon and A.~Shomer,
  {\it Superstrings on AdS(3) and symmetric products},
  JHEP {\bf 0012}, 003 (2000)
  {\tt [arXiv:hep-th/0009242]}.

\bibitem{Satoh}
  Y.~Satoh,
  {\it Three point functions and operator product expansion in the SL(2) conformal field theory},
  Nucl.\ Phys.\  {\bf B629}, 188-208 (2002).
  {\tt [arXiv:hep-th/0109059]}.

\bibitem{Ribault}
  S.~Ribault,
  {\it Knizhnik-Zamolodchikov equations and spectral flow in AdS(3) string theory},
  JHEP {\bf 0509}, 045 (2005).
  {\tt [arXiv:hep-th/0507114]}.

\bibitem{Baron}
  W.~H.~Baron, C.~A.~Nunez,
  {\it Fusion rules and four-point functions in the SL(2,R) WZNW model},
  Phys.\ Rev.\  {\bf D79}, 086004 (2009).
  {\tt [arXiv:0810.2768 [hep-th]]}.
  
\bibitem{Fjelstad}
  J.~Fjelstad,
  {\it On duality and extended chiral symmetry in the SL(2,R) WZW model},
  J.\ Phys.\ A {\bf A44}, 235404 (2011).
  {\tt [arXiv:1102.4196 [hep-th]]}.

\bibitem{GKS}
  A.~Giveon, D.~Kutasov and N.~Seiberg,
  {\it Comments on string theory on AdS(3)},
  Adv.\ Theor.\ Math.\ Phys.\  {\bf 2}, 733 (1998)
  {\tt [arXiv:hep-th/9806194]}; 
  D.~Kutasov and N.~Seiberg,
  {\it More comments on string theory on AdS(3)},
  JHEP {\bf 9904}, 008 (1999)
  {\tt [arXiv:hep-th/9903219]}.

\bibitem{deBoer:1998pp}
  J.~de Boer, H.~Ooguri, H.~Robins and J.~Tannenhauser,
  {\it String theory on AdS(3),}
  JHEP {\bf 9812}, 026 (1998)
  {\tt [arXiv:hep-th/9812046]}.

\bibitem{Teschner1999}
  J.~Teschner,
  {\it Operator product expansion and factorization in the $H_3^+$ WZNW
   model}, 
  Nucl.\ Phys.\ B {\bf 571}, 555 (2000)
  {\tt [arXiv:hep-th/9906215]}; 
  J.~Teschner,
  {\it On structure constants and fusion rules in the SL(2,C)/SU(2) WZNW
   model}, 
  Nucl.\ Phys.\ B {\bf 546}, 390 (1999)
  {\tt [arXiv:hep-th/9712256]}.



\bibitem{Zamolodchikov}
  A.~B.~Zamolodchikov and V.~A.~Fateev,
  {\it Operator Algebra and Correlation Functions in the Two-Dimensional
  Wess-Zumino SU(2) x SU(2) Chiral Model},
  Sov.\ J.\ Nucl.\ Phys.\  {\bf 43} (1986) 657
  [Yad.\ Fiz.\  {\bf 43} (1986) 1031].

\bibitem{Dotsenko}
  V.S.~Dotsenko,
  {\it The free field representation of the SU(2) conformal field theory},
  Nucl.\ Phys.\ B {\bf 338}, 747 (1990);
  V.~S.~Dotsenko,
  {\it Solving the SU(2) conformal field theory with the Wakimoto free
   field representation},
  Nucl.\ Phys.\ B {\bf 358}, 547 (1991).

\bibitem{Zarembo}
  K.~Zarembo,
  {\it Holographic three-point functions of semiclassical states},
  JHEP {\bf 1009}, 030 (2010)
  {\tt [arXiv:1008.1059 [hep-th]]}.

\bibitem{Costa:2010rz}
  M.~S.~Costa, R.~Monteiro, J.~E.~Santos and D.~Zoakos,
  {\it On three-point correlation functions in the gauge/gravity duality},
  JHEP {\bf 1011}, 141 (2010)
  {\tt [arXiv:1008.1070 [hep-th]]}.


\bibitem{Tseytlin}
  R.~Roiban and A.~A.~Tseytlin,
  {\it On semiclassical computation of 3-point functions of closed string vertex
  operators in $AdS_5 x S^5$},
  Phys.\ Rev.\  D {\bf 82}, 106011 (2010)
  {\tt [arXiv:1008.4921 [hep-th]]]}.

\bibitem{Tseytlin2}
  E.~I.~Buchbinder and A.~A.~Tseytlin,
  {\it Semiclassical four-point functions in $AdS_5 \times S^5$},
  JHEP {\bf 1102}, 072 (2011)
  {\tt [arXiv:1012.3740 [hep-th]]]}.
  
\bibitem{Tseytlin3}
  J.~G.~Russo and A.~A.~Tseytlin,
  {\it Large spin expansion of semiclassical 3-point correlators in $AdS_5 \times S^5$},
  JHEP {\bf 1102}, 029 (2011)
  {\tt [arXiv:1012.2760 [hep-th]]]}.
 
\bibitem{Escobedo}
  J.~Escobedo, N.~Gromov, A.~Sever, P.~Vieira,
  {\it Tailoring Three-Point Functions and Integrability II. Weak/strong coupling match},
   {\tt [arXiv:1104.5501 [hep-th]]}.

 
\bibitem{Troost}
  S.~K.~Ashok, R.~Benichou and J.~Troost,
  {\it Asymptotic Symmetries of String Theory on $AdS_3 \times S^3$ with Ramond-Ramond
  Fluxes},
  JHEP {\bf 0910}, 051 (2009)
  {\tt [arXiv:0907.1242 [hep-th]]}.


\end{thebibliography}
\end{document}